\documentclass[english, floatfix, amsfonts, amsmath, amssymb, a4paper, showpacs,
               twocolumn]{revtex4}

\usepackage[T1]{fontenc}
\usepackage[latin9]{inputenc}
\usepackage{babel}
\usepackage{graphicx}
\usepackage{float}
\makeatletter
\@ifundefined{showcaptionsetup}{}{%
 \PassOptionsToPackage{caption=false}{subfig}}
\usepackage{subfig}
\makeatother

\begin{document}


\title{Entanglement in a molecular three-qubit system}
\author{Amit Kumar Pal and Indrani Bose}
\email{indrani@bosemain.boseinst.ac.in}
\affiliation{%
Department of Physics, Bose Institute, 93/1, A. P. C. Road, Kolkata - 700009\\
}
\date{\today}


\begin{abstract}
We study the entanglement properties of a molecular three-qubit system described by the 
Heisenberg spin Hamiltonian with anisotropic exchange interactions and including an external 
magnetic field. The system exhibits first order quantum phase transitions by tuning two 
parameters, $x$ and $y$, of the Hamiltonian to specific values. The three-qubit chain is 
open ended so that there are two types of pairwise entanglement : nearest-neighbour (nn) and 
next-nearest-neighbour (nnn). We calculate the ground and thermal state concurrences, 
quantifying pairwise entanglement, as a function of the parameters $x$, $y$ and the 
temperature $T$. The entanglement threshold and gap temperatures are also determined as a 
function of the anisotropy parameter $x$. The results obtained are of relevance in understanding the 
entanglement features of the recently engineered molecular $Cr_{7}Ni$-$Cu^{2+}$-$Cr_{7}Ni$
complex which serves as a three-qubit system at sufficiently low temperatures.     
\end{abstract}


\pacs{03.67.Mn, 03.67.Bg, 03.65.Ud, 64.70.Tg, 75.10.Dg}


\maketitle


\section{Introduction}
Entanglement is a unique feature of quantum mechanical systems with no 
classical analogue. In an entangled state, two or more quantum particles 
have joint properties in the form of non-local correlations rather than 
individual identities. Entanglement is known to be a key resource in quantum 
information processing (QIP) tasks such as quantum computation, teleportation
and cryptography \cite{key-1}. Implementation of QIP protocols requires the assembly of
multi-qubit systems with the potential for generating controlled entanglement.
Natural examples of qubits, which are two-level systems, include spin-$\frac{1}{2}$
particles, photons with two states of polarization and trapped ions with two 
atomic states. In recent years, molecular nanomagnets have been proposed as 
appropriate candidates for qubit encoding and manipulation \cite{key-2,key-3}. 
A specific example is provided by antiferromagnetic (AFM) $Cr_{7}Ni$ rings which reduce 
to effective spin-$\frac{1}{2}$ systems at low temperatures. Each octagonal ring 
consists of one $Ni^{2+}$ and seven $Cr^{3+}$ ions with AFM coupling between 
neighbouring ions. A variety of experimental techniques have been used to 
characterize the rings. The rings have spin-$\frac{1}{2}$ ground states and 
behave as qubits at sufficiently low temperatures as the excited-state 
multiplets remain unoccupied. Also the rings have been demonstrated to possess
long decoherence times, an ideal requirement for several QIP tasks.

\begin{figure}
\begin{center}
\includegraphics[scale=0.45]{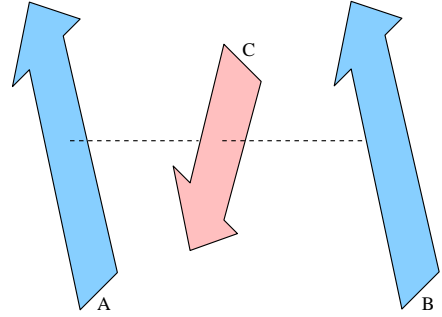}
\end{center}
\caption{A molecular three-qubit system in which the qubits $A$ and $B$ represent two 
$Cr_{7}Ni$ rings (see figure 1(a) of reference \cite{key-3}) and qubit $C$ represents the bridging
ion $Cu^{2+}$. The rings and the ion are effective spin-$\frac{1}{2}$ systems represented by 
solid arrows.}
\end{figure}

Recently, Timco \textit{et al} \cite{key-3} have engineered a coherent coupling between two $Cr_{7}Ni$
rings, serving as molecular spin qubits, via a central $Cu^{2+}$ ion which acts 
as a third qubit. The $Cr_{7}Ni$-$Cu^{2+}$-$Cr_{7}Ni$ complex is equivalent to a 
three qubit system with the $Cu^{2+}$ ion serving as a ``linker''. The coupling 
between the spins of the rings is tunable by a proper choice of the linker.
In a microscopic approach, the spin Hamiltonian describing the three-qubit system
can be written as \cite{key-3}
\begin{equation}
 H=H^{A}+H^{B}+H^{C}+H^{AC}+H^{BC}
\end{equation}
where the labels $A,B$ and $C$ correspond to the two rings and the magnetic linker 
respectively. The terms $H^{A}$ and $H^{B}$ individually describe the $Cr_{7}Ni$ 
rings:
\begin{eqnarray}
& & H^{A}=H^{B} =  \sum_{i=1}^{8}J_{i}\overrightarrow{S}_{i}.\overrightarrow{S}_{i+1} \nonumber \\
& & +\sum_{i=1}^{8}d_{i}S_{i,z}^2 + H_{dip} +\mu_{B}\overrightarrow{B}.\sum_{i=1}^{8}
\overrightarrow{\overrightarrow{g_{i}}}.\overrightarrow{S}_{i}
\end{eqnarray}
with $z$ along the ring axis. The successive terms in the Hamiltonian correspond
to isotropic exchange ($J_{i}$), axial crystal field ($d_{i}$), dipole-dipole 
couplings ($H_{dip}$) between eight individual spins $\overrightarrow{S}_{i}$ and 
the Zeeman coupling to the magnetic field $\overrightarrow{B}$ with 
$\overrightarrow{\overrightarrow{g_{i}}}$ being the gyromagnetic tensor. 
The term $H^{C}$ in equation (1) is
\begin{equation}
 H^{C}=\overrightarrow{B}.\overrightarrow{\overrightarrow{g}}_{Cu}.
 \overrightarrow{S}_{Cu}
\end{equation}
whereas the terms $H^{AC}/H^{BC}$ are:
\begin{equation}
 H^{AC}=H^{BC}=J^{\prime}\overrightarrow{S}_{Cu}.\left(
 \overrightarrow{S}_{Cr}+\overrightarrow{S}_{Ni}\right)
\end{equation}
where the spins $\overrightarrow{S}_{Cr}$ and $\overrightarrow{S}_{Ni}$ correspond
to $\overrightarrow{S}_{1}$ and $\overrightarrow{S}_{8}$ in their respective rings
as these spins are located on the edge of the octagon bound to the $Cu$ link.
Since $J^{\prime}<<J_{i}$'s, the intra-ring exchange constants, the low-temperature 
behaviour of the $Cr_{7}Ni$-$Cu^{2+}$-$Cr_{7}Ni$  complex is determined by the splitting
of the lowest eight energy levels. The behaviour can be reproduced in terms of an 
effective three-spin Hamiltonian \cite{key-3}:
\begin{eqnarray}
& & H = \bar{J}\sum_{i=A,B}\vec{S}_{i}.\vec{S}_{C}+\mu_{B}\overrightarrow{B}.\sum_{i=A,B,C}
\overrightarrow{\overrightarrow{g_{i}}}.\overrightarrow{S}_{i} \nonumber \\
& & +D_{ex}\sum_{i=A,B}(2S_{i,z}S_{C,z}-S_{i,x}S_{C,x}-S_{i,y}S_{C,y})
\end{eqnarray}
where $\overrightarrow{S}_{A,B,C}$ represent spin-$\frac{1}{2}$ operators, $\bar{J}$ 
is the strength of the effective $Cu$-ring isotropic exchange, $\overrightarrow{
\overrightarrow{g}}_{A,B}$ are the $g$-tensors of the ring ground doublet, 
$\overrightarrow{\overrightarrow{g}}_{C}=\overrightarrow{\overrightarrow{g}}_{Cu}$,
and $D_{ex}$ is an effective $Cu$-ring axial exchange originating from the projection
of the rings' dipolar and crystal-field anisotropies. Equation (5) represents the 
$Cr_{7}Ni$-$Cu^{2+}$-$Cr_{7}Ni$ system as a linear chain of three coupled qubits with open
boundaries. The three-qubit system has ground and thermal states which are entangled.
One can focus on two types of entanglement : pairwise, i.e. between two qubits and 
three-party entanglement involving all the three qubits. The Greenberger-Horne-Zeilinger 
(GHZ) and Werner (W) states \cite{key-4,key-5,key-6} defined as
\begin{equation}
\begin{array}{c}
  |GHZ\rangle=\frac{1}{\sqrt{2}}\left(|\uparrow\uparrow\uparrow\rangle
  +|\downarrow\downarrow\downarrow\rangle\right)\\
  |W\rangle=\frac{1}{\sqrt{3}}\left(|\uparrow\uparrow\downarrow\rangle+
  |\uparrow\downarrow\uparrow\rangle+|\downarrow\uparrow\uparrow\rangle\right)
\end{array}
\end{equation}
represent two fundamentally non-equivalent entangled states of three qubits.
In the first case, the pairwise entanglement for all the qubit pairs is zero
and one has genuine three party entanglement known as the residual entanglement.
The nomenclature arises from the Coffman-Kundu-Wootters (CKW) inequality \cite{key-7} for
a three qubit system given by,
\begin{equation}
 \tau_{1}\geq \tau_{2}=\sum_{j\neq i}C_{ij}^{2}
\end{equation}
\noindent where $\tau_{1}$ represents the one-tangle corresponding to the entanglement 
between the $i$th qubit and the rest of the system and $C_{ij}^{2}$ is the 
square of concurrence, a measure of the entanglement between the $i$th and $j$th qubits. 
The one-tangle $\tau_{1}$ is determined as 
$\tau_{1}=4det\rho^{(1)}$ where $\rho^{(1)}$ is the single-site reduced 
density matrix. The residual entanglement is given by the difference between 
$\tau_{1}$ and $\tau_{2}$ and hence provides a measure of quantum correlations
which cannot be expressed in terms of pairwise correlations. The GHZ state 
has the maximum possible value of 1 for the three-party (residual) entanglement.
The W state, on the other hand, possesses only pairwise entanglement between
all qubit pairs and the magnitude of the residual entanglement is zero. 
Timco \textit{et al} \cite{key-3} have provided a prescription for the generation of GHZ and W 
states using a sequence of microwave pulses applied to the molecular three-qubit system.

In this paper, we study the entanglement properties of the ground and thermal
states of the molecular three-qubit system described by the reduced Hamiltonian
in equation (5). We specially focus on the variation of entanglement measures as a 
function of the parameters of the Hamiltonian. Wang \textit{et al} \cite{key-8} have earlier 
studied thermal entanglement in the three-qubit Heisenberg XXZ model. The Hamiltonian 
considered by them satisfies periodic boundary condition and includes anisotropic 
exchange interaction and magnetic field terms. The molecular three qubit system
considered in this paper has the structure of an open chain and the entanglement 
features turn out to be different from those of the three-qubit Heisenberg ring. 
The experimental demonstration that the coupling between the molecular spin clusters
can be controlled without disturbing the intra-cluster interactions provides the 
impetus for characterizing the entanglement properties of the molecular 
three-qubit system. 

 
\section{Ground State Entanglement} 
We consider the molecular three-qubit system to be in an external magnetic field
pointed in the $z$ direction. The Hamiltonian (equation (5)) then reduces to
\begin{eqnarray}
& & H = \bar{J}\sum_{i=A,B}\vec{S}_{i}.\vec{S}_{C}+g\mu_{B}B\sum_{i=A,B,C}S_{i}^{z} \nonumber \\
& & +D_{ex}\sum_{i=A,B}(2S_{i,z}S_{C,z}-S_{i,x}S_{C,x}-S_{i,y}S_{C,y})
\end{eqnarray}
This can be rewritten in the form,
\begin{eqnarray}
& & \bar{H}=H/\bar{J}=\left(1+2x\right)\left(S_{A,z}S_{C,z}+S_{B,z}S_{C,z}\right) \nonumber \\
& & +\frac{1}{2}\left(1-x\right)\left(S_{A}^{+}S_{C}^{-}+S_{A}^{-}S_{C}^{+}
+S_{B}^{+}S_{C}^{-}+S_{B}^{-}S_{C}^{+}\right) \nonumber \\
& & +y\left(S_{A,z}+S_{B,z}+S_{C,z}\right)
\end{eqnarray}
where $x=D_{ex}/\bar{J}$, $y=g\mu_{B}B/\bar{J}$ and $S^{+},S^{-}$ 
are the raising and lowering operators. Since the $z$-component of the total spin, 
$S_{z}^{tot}$, is a conserved quantity, the eigenvalue problem can be solved in the
separate subspaces corresponding to the different values of $S_{z}^{tot}$. The 
eigenvalues and the eigenstates are given by,

\bigskip{}
\underline{$S_{z}^{tot}=+\frac{3}{2}$}
\bigskip{}
\begin{equation}
 \begin{array}{c}
  |\psi_{1}\rangle=|\uparrow\uparrow\uparrow\rangle\\
  \medskip{}
  E_{1}=\frac{1}{2}\left(1+2x+3y\right)
 \end{array}
\end{equation}

\bigskip{}
\underline{$S_{z}^{tot}=+\frac{1}{2}$}
\bigskip{}
\begin{equation}
 \begin{array}{c}
  |\psi_{2}\rangle=\frac{1}{\sqrt{2}}\left(-|\uparrow\uparrow\downarrow\rangle
  +|\downarrow\uparrow\uparrow\rangle\right)\\
  E_{2}=\frac{y}{2}
 \end{array}
\end{equation}
\bigskip{}
\begin{equation}
 \begin{array}{c}
  |\psi_{3}\rangle=\frac{1}{A(x)}\left(|\uparrow\uparrow\downarrow\rangle
  -R(x)|\uparrow\downarrow\uparrow\rangle
  +|\downarrow\uparrow\uparrow\rangle\right)\\
  E_{3}=\frac{1}{4}\left\{\ 2y-U_{+}(x)\right\}\
 \end{array}
\end{equation}
\bigskip{}
\begin{equation}
 \begin{array}{c}
  |\psi_{4}\rangle=\frac{1}{B(x)}\left(|\uparrow\uparrow\downarrow\rangle
  -S(x)|\uparrow\downarrow\uparrow\rangle
  +|\downarrow\uparrow\uparrow\rangle\right)\\
  E_{4}=\frac{1}{4}\left\{\ 2y-U_{-}(x)\right\}\
 \end{array}
\end{equation}

\bigskip{}
\underline{$S_{z}^{tot}=-\frac{1}{2}$}
\bigskip{}
\begin{equation}
 \begin{array}{c}
  |\psi_{5}\rangle=\frac{1}{A(x)}\left(|\downarrow\downarrow\uparrow\rangle
  -R(x)|\downarrow\uparrow\downarrow\rangle
  +|\uparrow\downarrow\downarrow\rangle\right)\\
  E_{5}=\frac{1}{4}\left\{\ -2y-U_{+}(x)\right\}\
 \end{array}
\end{equation}
\bigskip{}
\begin{equation}
 \begin{array}{c}
  |\psi_{6}\rangle=\frac{1}{B(x)}\left(|\downarrow\downarrow\uparrow\rangle
  -S(x)|\downarrow\uparrow\downarrow\rangle
  +|\uparrow\downarrow\downarrow\rangle\right)\\
  E_{6}=\frac{1}{4}\left\{\ -2y-U_{-}(x)\right\}\
 \end{array}
\end{equation}
\bigskip{}
\begin{equation}
 \begin{array}{c}
  |\psi_{7}\rangle=\frac{1}{\sqrt{2}}\left(-|\downarrow\downarrow\uparrow\rangle
  +|\uparrow\downarrow\downarrow\rangle\right)\\
  E_{7}=-\frac{y}{2}
 \end{array}
\end{equation}

\bigskip{}
\underline{$S_{z}^{tot}=-\frac{3}{2}$}
\bigskip{}
\begin{equation}
 \begin{array}{c}
  |\psi_{8}\rangle=|\downarrow\downarrow\downarrow\rangle\\
  E_{8}=\frac{1}{2}\left(1+2x-3y\right)
 \end{array}
\end{equation}
In the above equations,
\begin{equation}
U_{\pm}(x)=\left\{\ 1+2x\pm\sqrt{3\left(4x^{2}-4x+3\right)}\right\}\
\end{equation}
\begin{equation}
 R(x)=\frac{-U_{+}(x)}{2(-1+x)}
\end{equation}
\begin{equation}
 S(x)=\frac{-U_{-}(x)}{2(-1+x)}
\end{equation}
\begin{equation}
 A(x)=\left[2+\left\{\frac{-U_{+}(x)}{2(-1+x)}\right\}^{2}
 \right]^{\frac{1}{2}}
\end{equation}
\begin{equation}
 B(x)=\left[2+\left\{\frac{-U_{-}(x)}{2(-1+x)}\right\}^{2}
 \right]^{\frac{1}{2}}
\end{equation}

We first consider the case of zero magnetic field (y=0). The eigenvalues then become
\begin{equation}
 \begin{array}{c}
  E_{1}=E_{8}=\frac{1+2x}{2}\\
  E_{2}=E_{7}=0\\
  E_{3}=E_{5}=\frac{-U_{+}(x)}{4}\\
  E_{4}=E_{6}=\frac{-U_{-}(x)}{4}
 \end{array}
\end{equation}

\begin{figure}
\begin{center}
\includegraphics[scale=0.7]{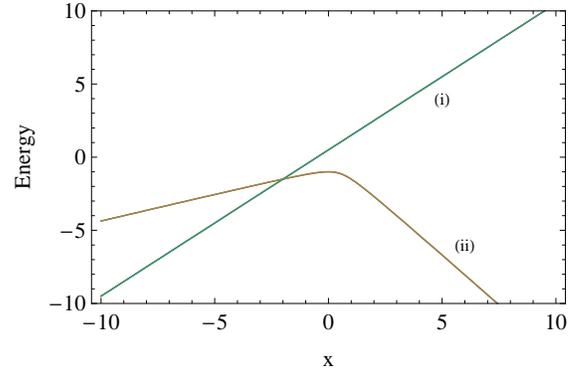}
\caption{Two lowest energy levels (i) $E_{1}$ and $E_{8}$ and (ii) $E_{3}$ and $E_{5}$ 
of the Hamiltonian (equation (9)) versus the parameter $x$ for $y=0$.}
\end{center}
\end{figure}
\noindent We assume $x$ to range over both positive and negative values. Figure 2 shows a plot of
the two lowest energy levels of the Hamiltonian (equation (9)) versus the parameter 
$x$. Each energy level is doubly-degenerate. The nature of the ground states change at 
$x=-2$, bringing about a 
first-order quantum phase transition (QPT). When $x$ is $<-2$, the ground states are the 
separable states $|\psi_{1}\rangle$ and $|\psi_{8}\rangle$. When $x$ is $>-2$, 
the doubly-degenerate ground state is described by the wave functions $|\psi_{3}\rangle$ 
and $|\psi_{5}\rangle$. At $x=1$, however, the Hamiltonian (9) becomes 
Ising-like, i.e., loses its quantum character and the degenerate ground states, 
$|\uparrow\downarrow\uparrow\rangle$ and $|\downarrow\uparrow\downarrow\rangle$, 
are separable. We now discuss the entanglement properties of the ground 
states. Because of the degeneracy, the ground state density matrix describes 
a mixed state with
\begin{equation}
 \rho=\frac{1}{2}\left(|\psi_{3}\rangle\langle\psi_{3}|
 +|\psi_{5}\rangle\langle\psi_{5}|\right)
\end{equation}
The reduced density matrix $\rho_{ij}$, $(i,j=A,B,C)$ is obtained from $\rho$ by 
tracing out the spin degrees of freedom associated with the spins which are not 
located at the sites $i$ and $j$. The reduced density matrix in the standard 
basis, $\left\{|\uparrow\uparrow\rangle,|\uparrow\downarrow\rangle,|\downarrow
\uparrow\rangle,|\downarrow\downarrow\rangle\right\}$, has the structure
\begin{equation}
 \left(
 \begin{array}{cccc}
  u & 0 & 0 & 0\\
  0 & w_{1} & y^{\star} & 0\\
  0 & y & w_{2} & 0\\
  0 & 0 & 0 & v
 \end{array}
 \right)
\end{equation}
The concurrence $C_{ij}$, a measure of the entanglement between a pair of spins
at sites $i$ and $j$, is given by \cite{key-9,key-10},
\begin{equation}
C_{ij}=2\;max\left(0,\left|y\right|-\sqrt{uv}\right)
\end{equation}
Figures 3(a) and 3(b) show the variation of $C_{AC}$ and $C_{AB}$ versus $x$. The 
analytical expressions for the concurrences are;
\begin{equation}
\begin{array}{c}
C_{AC}=C_{BC}=2\;max\left(0,\left|\frac{R}{A^{2}}\right|-\frac{1}{2A^{2}}\right)\\
C_{AB}=2\;max \left(0,\left|\frac{1}{A^{2}}\right|-\frac{R^{2}}{2A^{2}}\right)
\end{array}
\end{equation}  
\begin{figure}
\begin{center}
\subfloat[]{\includegraphics[scale=0.7]{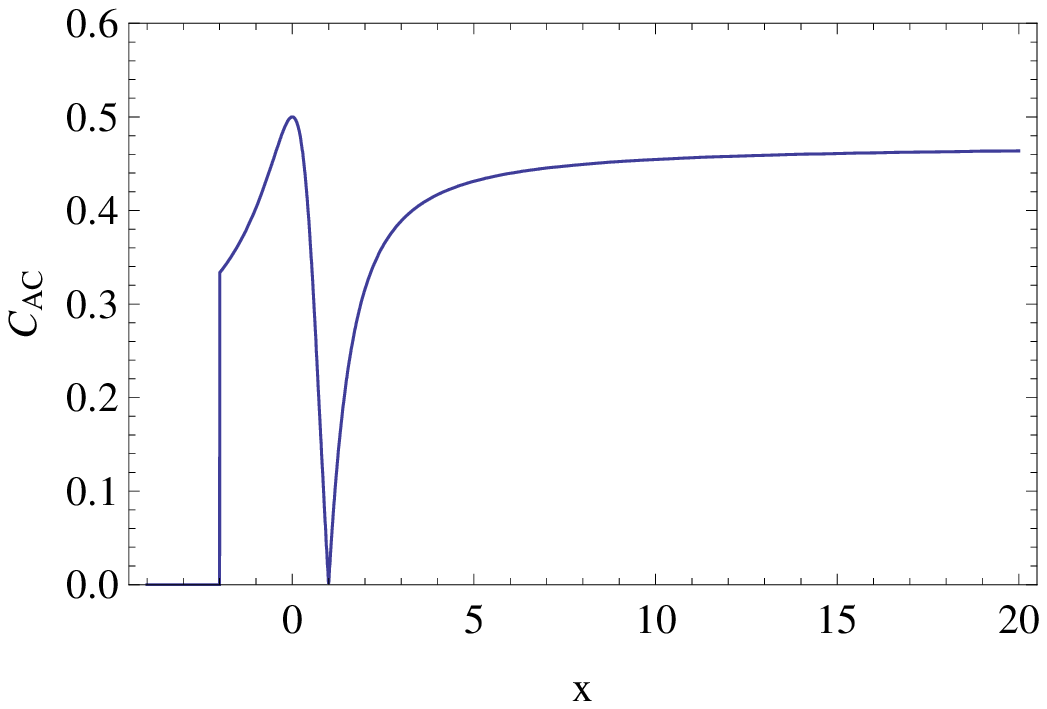}}
\smallskip{}
\subfloat[]{\includegraphics[scale=0.7]{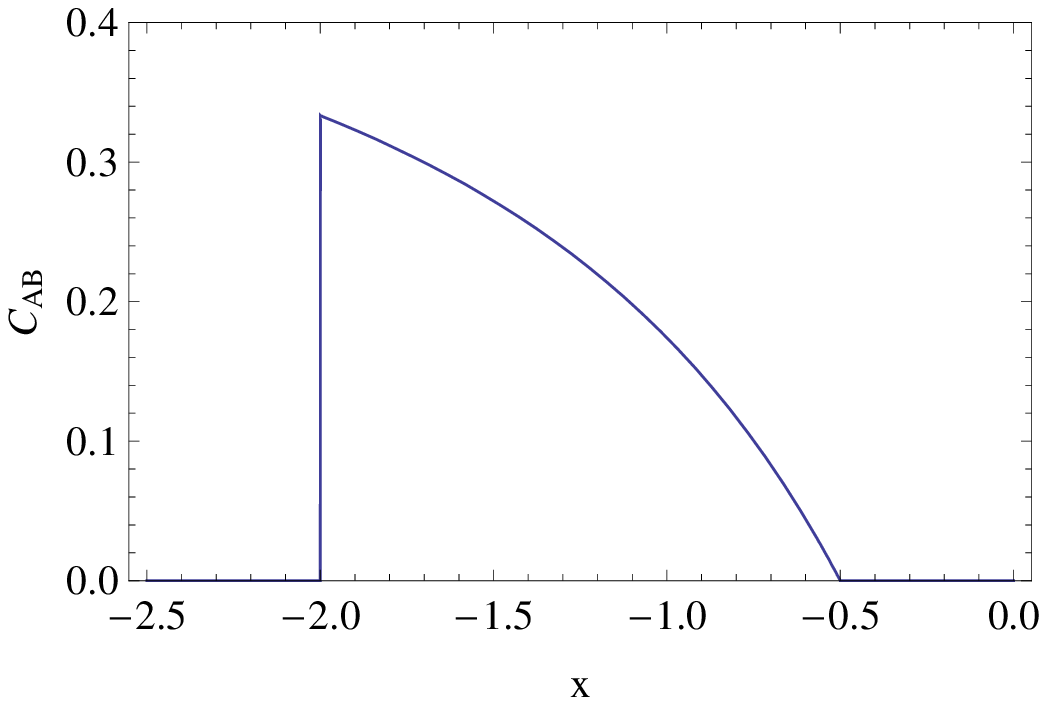}}
\caption{Variation of concurrences (a) $C_{AC}$ and (b) $C_{AB}$ versus x for $y = 0$.}
\end{center}
\end{figure}    
\noindent The  variation of $C_{BC}$ as a function of $x$ is identical with that of 
$C_{AC}$. We remind ourselves that $A$ and $B$ are the boundary spins and 
$C$ the central spin. A jump in the magnitude of the concurrence indicates 
a first order QPT \cite{key-11,key-12,key-13,key-14} which, as already mentioned, occurs
at $x=-2$. $C_{AC}$ and $C_{BC}$ both become zero at $x=1$ due to the  
separability of the ground state density matrix and then rises
as $x$ is increased to attain a saturation value $C_{AC}=\frac{1+2\sqrt{3}}{6+2\sqrt{3}}$ 
for large $x$. The entanglement between the boundary spins, however, has a non-zero value only
for negative values of $x$ and that too in a restricted range of $x$ values.
\begin{figure}
\begin{center}
\includegraphics[scale=0.7]{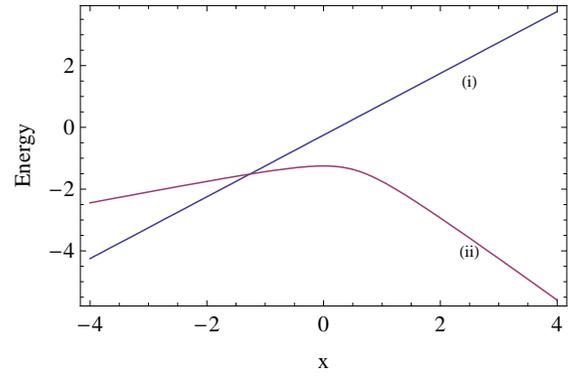}
\caption{Plot of the two lowest energies (i)$E_{8}$ and (ii)$E_{5}$ versus $x$ for $y = 0.5$.}
\end{center}
\end{figure}
\begin{figure}
\begin{center}
\includegraphics[scale=0.7]{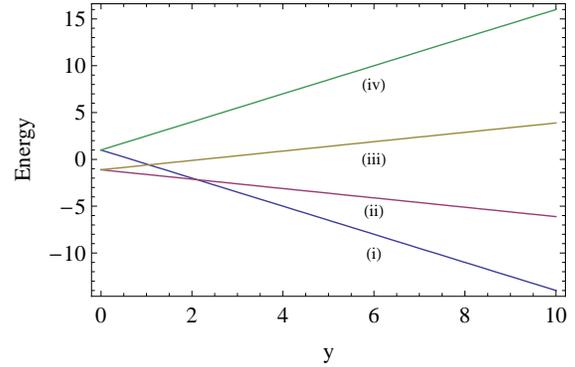}
\caption{Plot of the lowest energy levels (i)$E_{8}$, (ii)$E_{5}$, (iii)$E_{3}$ and (iv)$E_{1}$
as a function of $y$ for $x = 0.5$}
\end{center}
\end{figure}

We next consider the case of non-zero magnetic field ($y\neq 0$). Figure 4  
shows the plots of the two lowest energies, $E_{5}$ and $E_{8}$, versus $x$
for $y=0.5$. One finds that a first-order QPT occurs at 
a specific value of $x=x_{c}\left(=-\sqrt{\frac{1}{2}}-\sqrt{\frac{7}{12}}\right)$ 
indicating a change in the nature of the ground state. Figure 5 
shows the variation of $E_{5}$ and $E_{8}$ as a function of $y$ for $x=0.5$. 
Again, one notes the occurrence of a first-order QPT 
at a specific value of $y=y_{c}$. The external magnetic field removes the ground state 
degeneracy of the zero-field case and the three-qubit system has a unique ground 
state. Figures 6(a) and 6(b) show the variation of the concurrences $C_{AC}$ and 
$C_{AB}$ versus $x$ for $y=0.5$. In this case, the analytical expressions for the 
concurrences are;
\begin{equation}
\begin{array}{c}
C_{AC}=C_{BC}=2\;max\left(0,\left|\frac{R}{A^{2}}\right|\right)\\
C_{AB}=2\;max \left(0,\left|\frac{1}{A^{2}}\right|\right)
\end{array}
\end{equation}
\begin{figure}
\begin{center}
\subfloat[]{\includegraphics[scale=0.7]{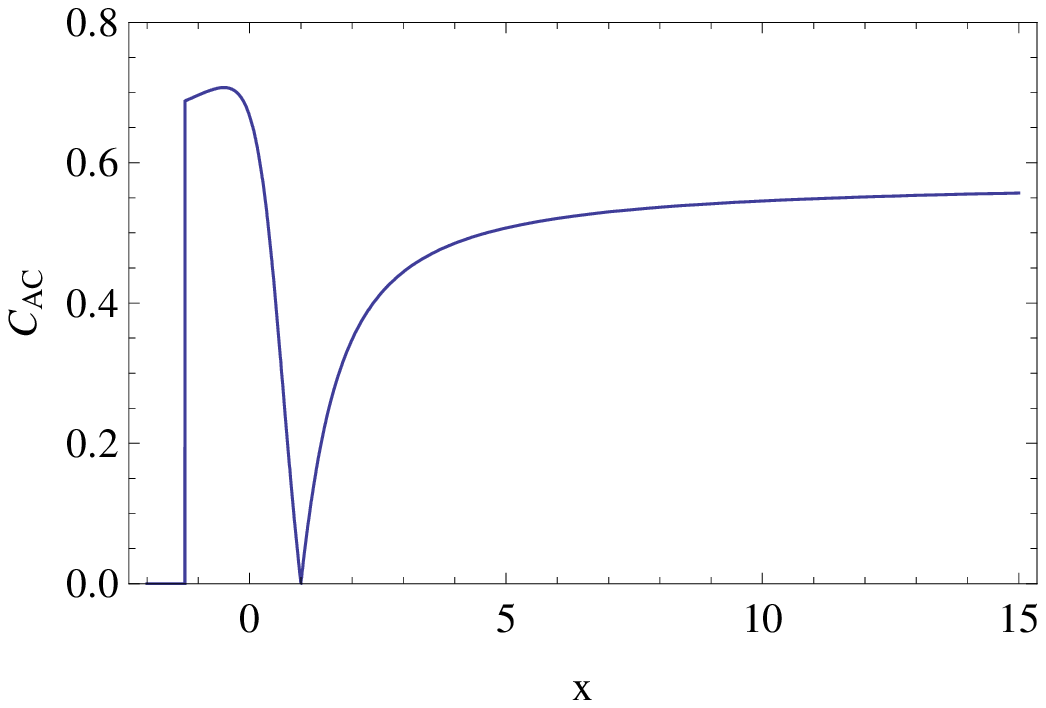}}
\smallskip{}
\subfloat[]{\includegraphics[scale=0.7]{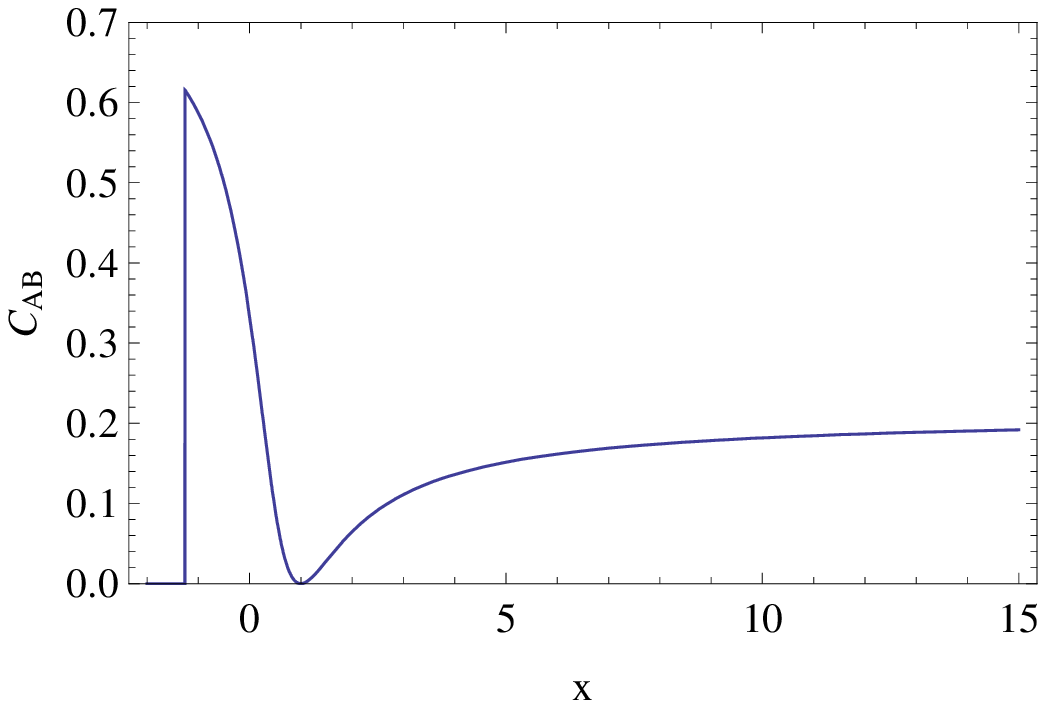}}
\caption{Variation of concurrences (a) $C_{AC}$ and (b) $C_{AB}$ versus $x$ for $y = 0.5$.}
\end{center}
\end{figure}
\noindent In zero magnetic field $(y=0)$, the next-nearest-neighbour (nnn) concurrence $C_{AB}$ 
has non-zero values only in a restricted 
range of negative $x$ values (figure 3(b)) whereas in the presence of a magnetic field
$(y\neq 0)$, $C_{AB}$ is non-zero in a range of both negative and positive $x$ values. The 
magnitude of the nnn entanglement is less than that of the nn entanglement 
for both $y=0$ and $y\neq 0$. As $y$ increases, one finds that $x_{c}$, the first 
order QPT point shifts towards more positive values. For sufficiently high values of 
$y$, entanglement exists only for positive values of $x$. This is so provided $y$ is 
less than the critical value $y_{c}$ (which depends upon $x$) 
at which a first-order QPT takes place to a separable ground state.
Figures 7(a) and 7(b) show the plots of the nn and nnn 
concurrences, $C_{AC}(=C_{BC})$ and $C_{AB}$ respectively versus $y$ for $x=0.5$. 
The concurrences have constant values for $y<y_{c}$, the QPT point, and jump 
discontinuously to zero values at $y=y_{c}$. As $x$ increases, the value of $y_{c}$
also increases. In the case of non-zero magnetic field, $y\neq 0$, 
the ground state is non-degenerate and the density matrix represents a pure state. 
In this case, the one-tangle $\tau_{1}$(defined in equation (7)) can be calculated. For 
any choice of the central spin, the residual entanglement involving three spins is 
found to be zero so that only pairwise entanglement exists in the ground state. The 
ground states thus belong to the class of $W$ rather than $GHZ$ states. 
\begin{figure}
\begin{center}
\subfloat[]{\includegraphics[scale=0.7]{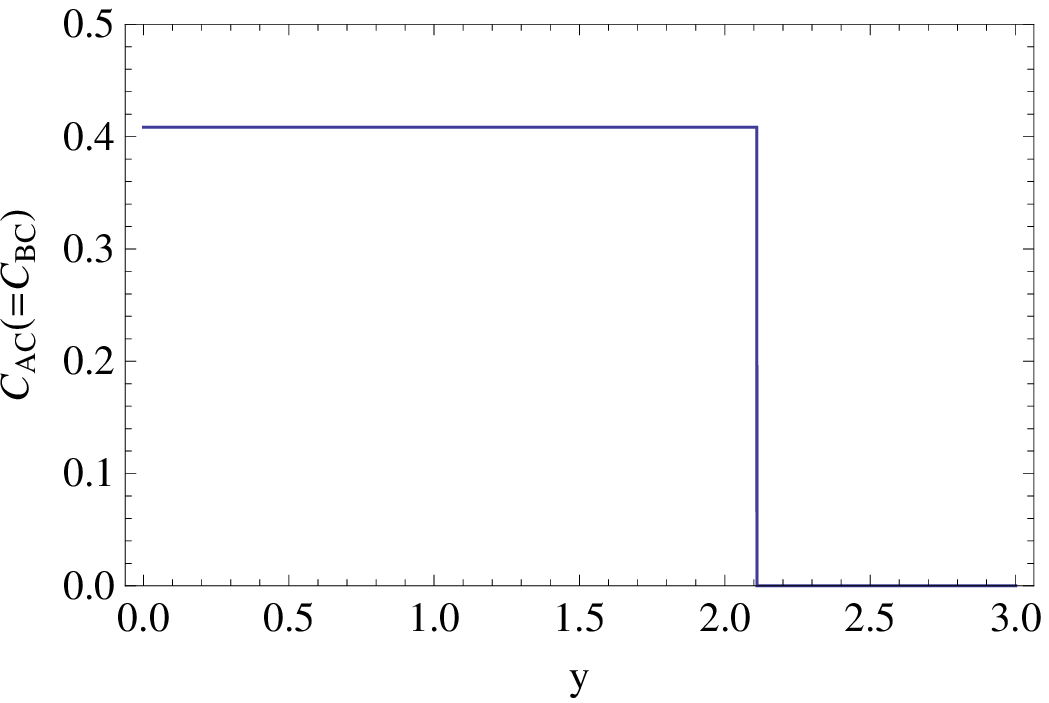}}
\smallskip{}
\subfloat[]{\includegraphics[scale=0.7]{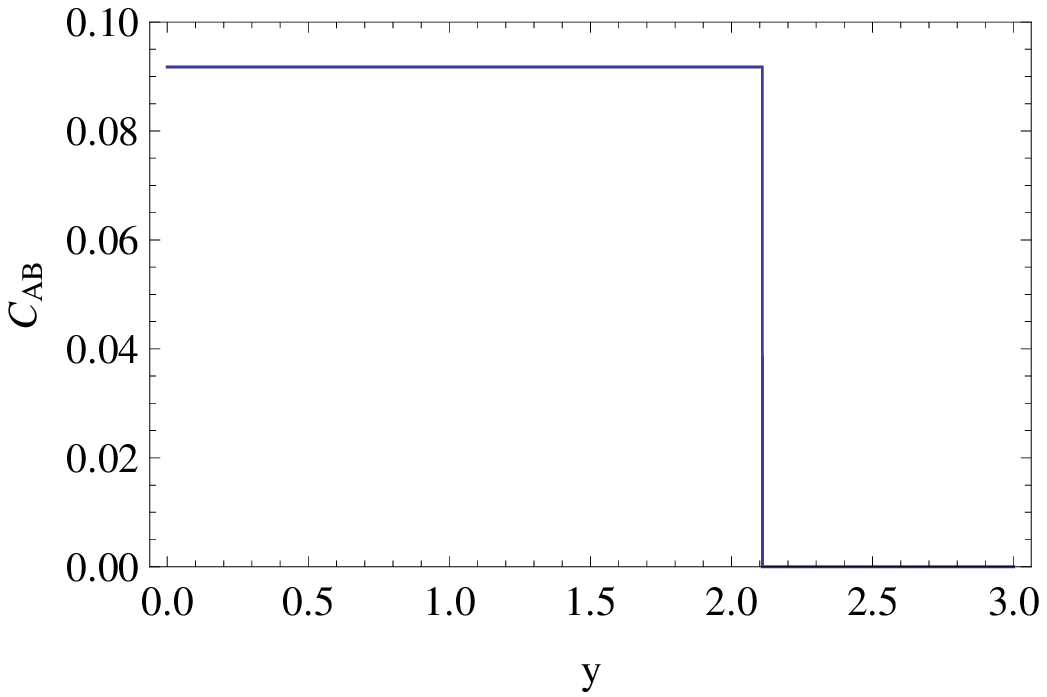}}
\caption{Variation of concurrences (a) $C_{AC}$ and (b) $C_{AB}$ versus $y$ for $x = 0.5$.}
\end{center}
\end{figure}


\section{Thermal state entanglement}
We next discuss the finite-temperature entanglement properties of the molecular 
three-qubit system. The thermal density matrix $\rho(T)=\frac{1}{Z}\exp{(-\beta H)}(\beta
=\frac{1}{T},k_{B}=1)$ now replaces the ground state density-matrix with $Z$ denoting 
the partition function of the system. The reduced density matrix $\rho_{ij}(T)$ has 
the same form as in equation (26) with $C_{ij}(T)$ given by \cite{key-15}
\begin{equation}
 C_{ij}(T)=\frac{2}{Z}\;max\left(0,\left|y(T)\right|-\sqrt{u(T)v(T)}\right)
\end{equation}
For the three-qubit system, the thermal density matrix is
\begin{equation}
 \rho(T)=\frac{1}{Z}\sum_{k=1}^{8}\exp{(-\beta E_{k})}
 |\psi_{k}\rangle\langle\psi_{k}|
\end{equation}
where the $|\psi_{k}\rangle$'s and $E_{k}$'s are given in Eqs. (10)-(17) for the 
general case $y\neq 0$. When $y=0$, the energy eigenvalues are as shown in 
equation (23). We first consider the case of zero magnetic field $(y=0)$. The matrix 
elements $u,v$ and $y$ of the reduced density matrix $\rho_{AC}(T)$ are
\begin{equation}
 u=v=\left(e^{\frac{-E_{1}}{T}}+\frac{1}{A^{2}}e^{\frac{-E_{3}}{T}}
 +\frac{1}{B^{2}}e^{\frac{-E_{4}}{T}}+\frac{1}{2}\right)
\end{equation}
\begin{equation}
 y=y^{\star}=\left(-\frac{2R}{A^{2}}e^{-\frac{E_{3}}{T}}
  -\frac{2S}{B^{2}}e^{-\frac{E_{4}}{T}}\right)
\end{equation}
The reduced density matrix $\rho_{BC}(T)$ has the same matrix elements as in 
equations (31) and (32). For the nnn concurrence, $C_{AB}$, the matrix elements of
the reduced density matrix are :
\begin{equation}
 u=v=\left(e^{\frac{-E_{1}}{T}}+\frac{R^{2}}{A^{2}}e^{\frac{-E_{3}}{T}}
 +\frac{S^{2}}{B^{2}}e^{\frac{-E_{4}}{T}}\right) 
\end{equation}
\begin{equation}
 y=y^{\star}=\left(\frac{2}{A^{2}}e^{-\frac{E_{3}}{T}}
  +\frac{2}{B^{2}}e^{-\frac{E_{4}}{T}}-1\right)
\end{equation}
\begin{figure}
\begin{center}
\subfloat[]{\includegraphics[scale=0.7]{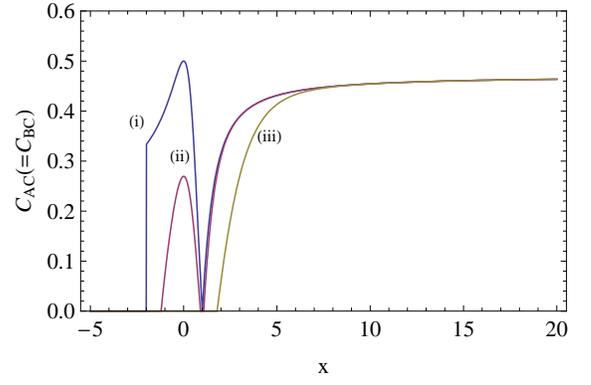}}
\smallskip{}
\subfloat[]{\includegraphics[scale=0.7]{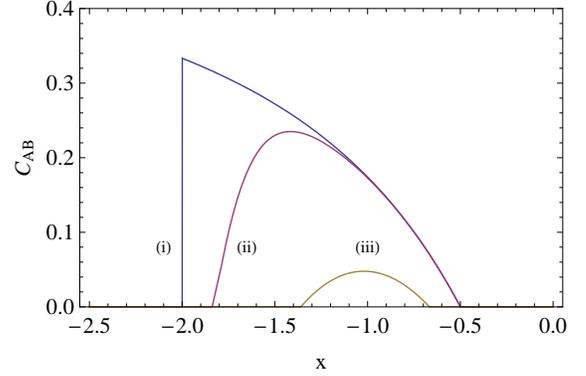}}
\caption{Variation of concurrences (a) $C_{AC}$ and (b) $C_{AB}$ versus $x$ at different 
temperatures $T$ with $y = 0$. The different temperature values are \textbf{(a)} (i) $T=0$, 
(ii) $T=0.5$, (iii) $T=1.5$ and \textbf{(b)} (i) $T=0$, (ii) $T=0.1$, (iii) $T=0.3$} 
\end{center}
\end{figure}
\noindent Figures 8(a) and 8(b) show the plots of $C_{AC}$ and $C_{AB}$ respectively as a 
function of $x$ for different values of the temperature $T$. As $T$ increases, the 
range of $x$ values for which $C_{AC}\neq 0$ shifts towards more positive values. 
Figures 9(a) and 9(b) show $C_{AC}$ and $C_{AB}$ versus $T$ for negative values of $x$. 
One can obtain similar plots for $C_{AC}$ when $x$ is $>0$. For both the nn and nnn
entanglements, one can define threshold temperatures $T_{C}^{(1)}$ and $T_{C}^{(2)}$  
respectively, beyond which the concurrences \cite{key-8,key-16} have zero values. 
Figures 10(a) and 10(b) show how $T_{C}^{(1)}$ and $T_{C}^{(2)}$ vary with $x$ for different 
values of $y$.
\begin{figure}
\begin{center}
\subfloat[]{\includegraphics[scale=0.7]{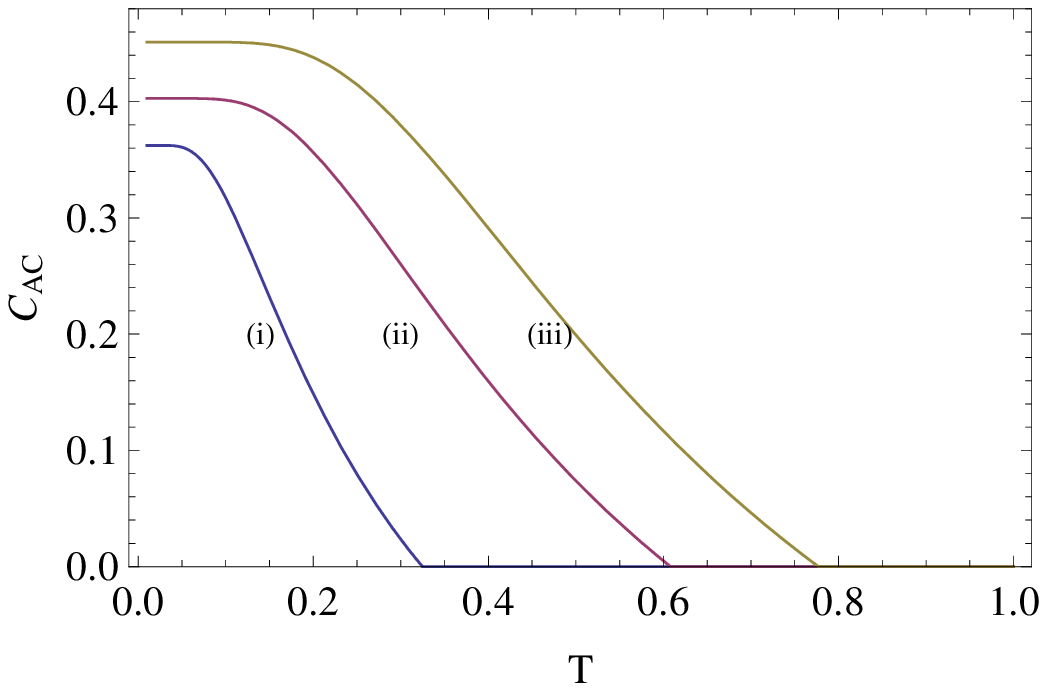}}
\smallskip{}
\subfloat[]{\includegraphics[scale=0.7]{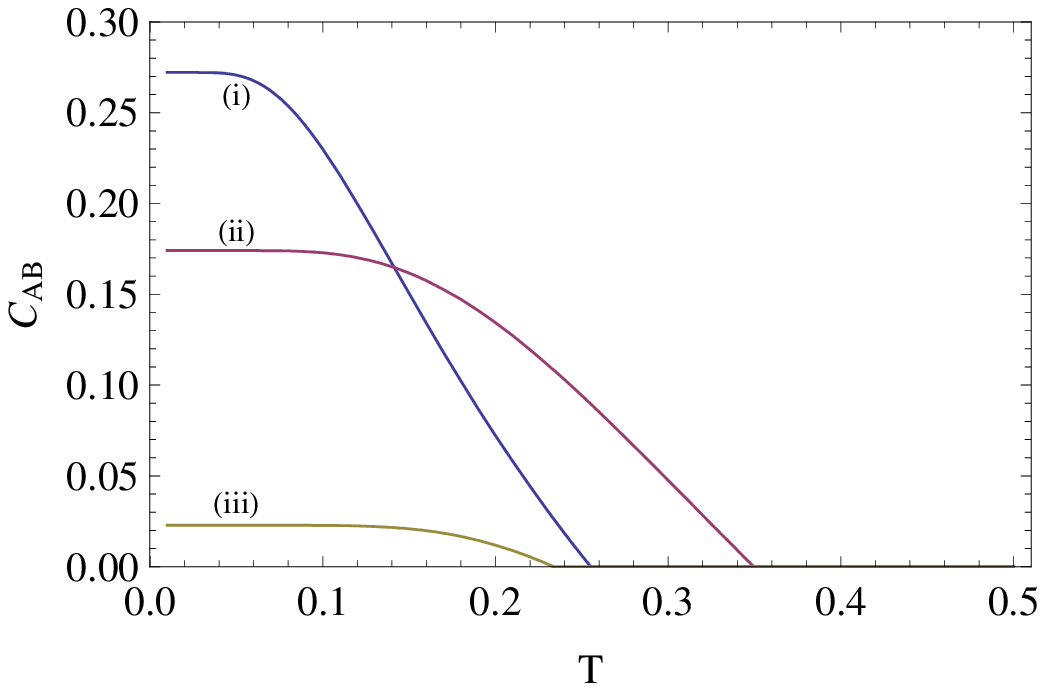}}
\caption{Variation of concurrences (a) $C_{AC}$ and (b) $C_{AB}$  versus $T$ for different 
negative values of $x$ with $y = 0$. The different $x$ values are (i) $x=-1.5$, (ii) $x=-1$, 
and (iii) $x=-0.55$ for both (a) and (b).}
\end{center}
\end{figure}
\begin{figure}
\begin{center}
\subfloat[]{\includegraphics[scale=0.7]{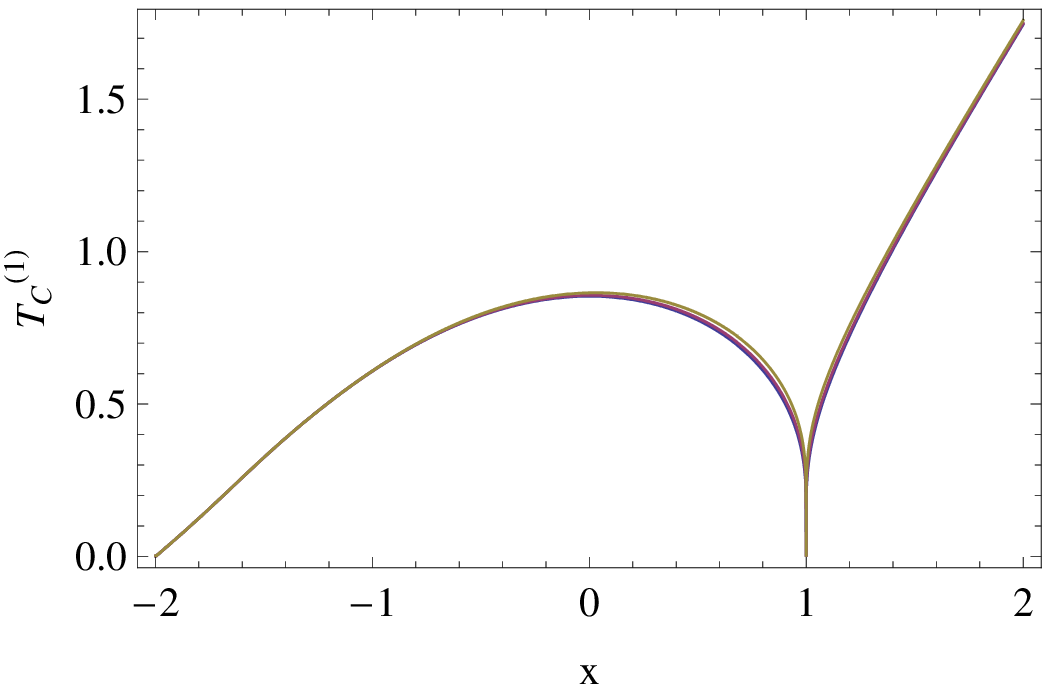}}
\smallskip{}
\subfloat[]{\includegraphics[scale=0.7]{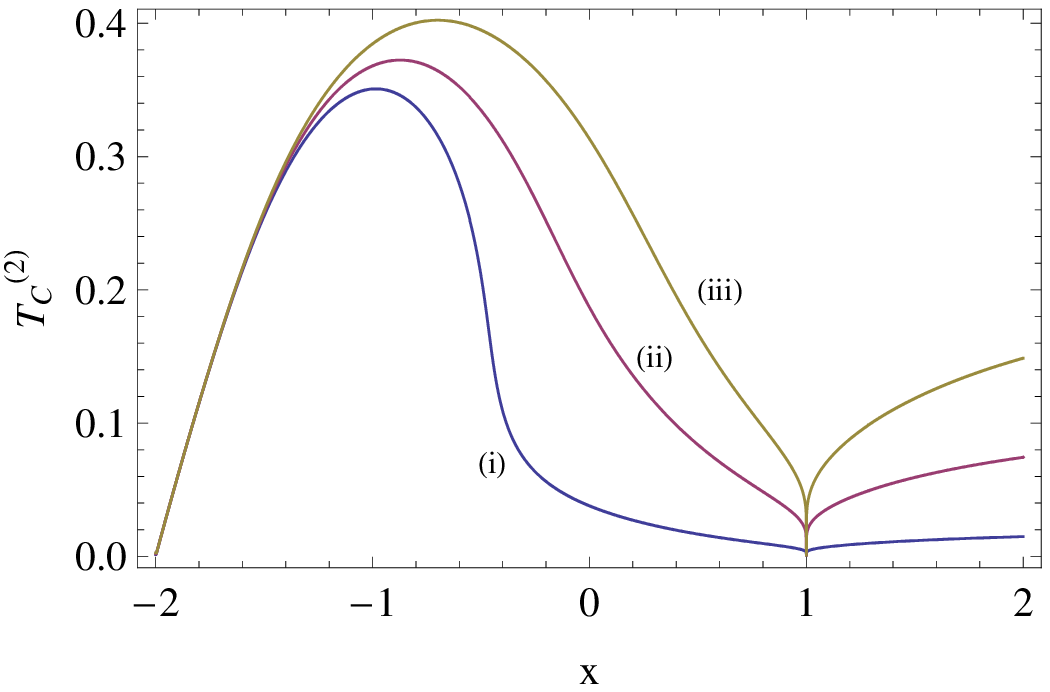}}
\caption{The threshold entanglement temperature (a) $T_{C}^{(1)}$ versus $x$ for nn 
entanglement and (b) $T_{C}^{(2)}$ versus $x$ for nnn entanglement with 
(i) $y=0.1$ (ii) $y=0.5$ and (iii) $y=1$. $T_{C}^{(1)}$ has a very weak dependence on the  
values of $y$.}
\end{center}
\end{figure}

For non-zero magnetic field, $y\neq 0$, the matrix elements of 
the reduced density matrix $\rho_{AC}(=\rho_{BC})$ are given by,
\begin{equation}
u=\left(e^{-\frac{E_{1}}{T}}+\frac{1}{2}e^{-\frac{E_{2}}{T}}
  +\frac{1}{A^{2}}e^{-\frac{E_{3}}{T}}+\frac{1}{B^{2}}e^{-\frac{E_{4}}{T}}\right) 
\end{equation}
\begin{equation}
v=\left(e^{-\frac{E_{8}}{T}}+\frac{1}{2}e^{-\frac{E_{7}}{T}}
  +\frac{1}{A^{2}}e^{-\frac{E_{5}}{T}}+\frac{1}{B^{2}}e^{-\frac{E_{6}}{T}}\right) 
\end{equation}
\begin{equation}
\begin{array}{c}
y=y^{\star}\\
=\left(-\frac{R}{A^{2}}e^{-\frac{E_{3}}{T}}-\frac{S}{B^{2}}e^{-\frac{E_{4}}{T}}
-\frac{R}{A^{2}}e^{-\frac{E_{5}}{T}}-\frac{S}{B^{2}}e^{-\frac{E_{6}}{T}}\right)
\end{array}
\end{equation}
The corresponding matrix elements for the nnn reduced density matrix are:
\begin{equation}
u=\left(e^{-\frac{E_{1}}{T}}+\frac{R^{2}}{A^{2}}e^{-\frac{E_{3}}{T}}
  +\frac{S^{2}}{B^{2}}e^{-\frac{E_{4}}{T}}\right)  
\end{equation}
\begin{equation}
v=\left(e^{-\frac{E_{8}}{T}}+\frac{R^{2}}{A^{2}}e^{-\frac{E_{5}}{T}}
  +\frac{S^{2}}{B^{2}}e^{-\frac{E_{6}}{T}}\right) 
\end{equation}
\begin{equation}
\begin{array}{c}
y=y^{\star}= -\frac{1}{2}\left(e^{-\frac{E_{2}}{T}}+e^{-\frac{E_{7}}{T}}\right)\\
+\frac{1}{A^{2}}\left(e^{-\frac{E_{3}}{T}}+e^{-\frac{E_{5}}{T}}\right)
+\frac{1}{B^{2}}\left(e^{-\frac{E_{4}}{T}}+e^{-\frac{E_{6}}{T}}\right)
\end{array}
\end{equation}
\begin{figure}
\begin{center}
\subfloat[]{\includegraphics[scale=0.7]{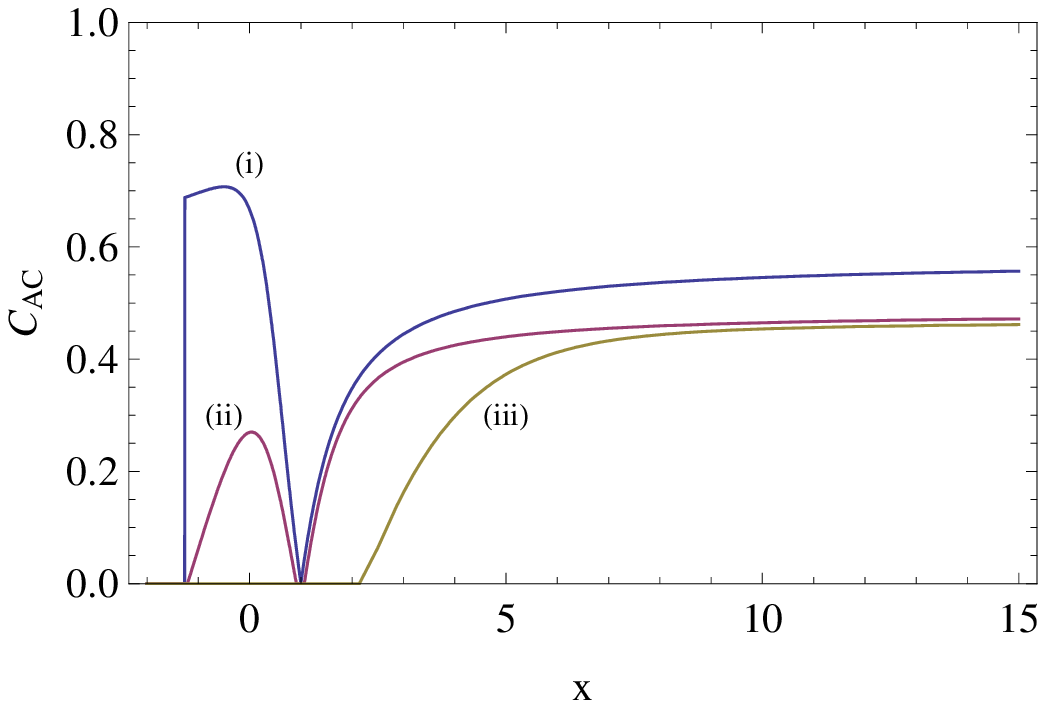}}
\smallskip{}
\subfloat[]{\includegraphics[scale=0.7]{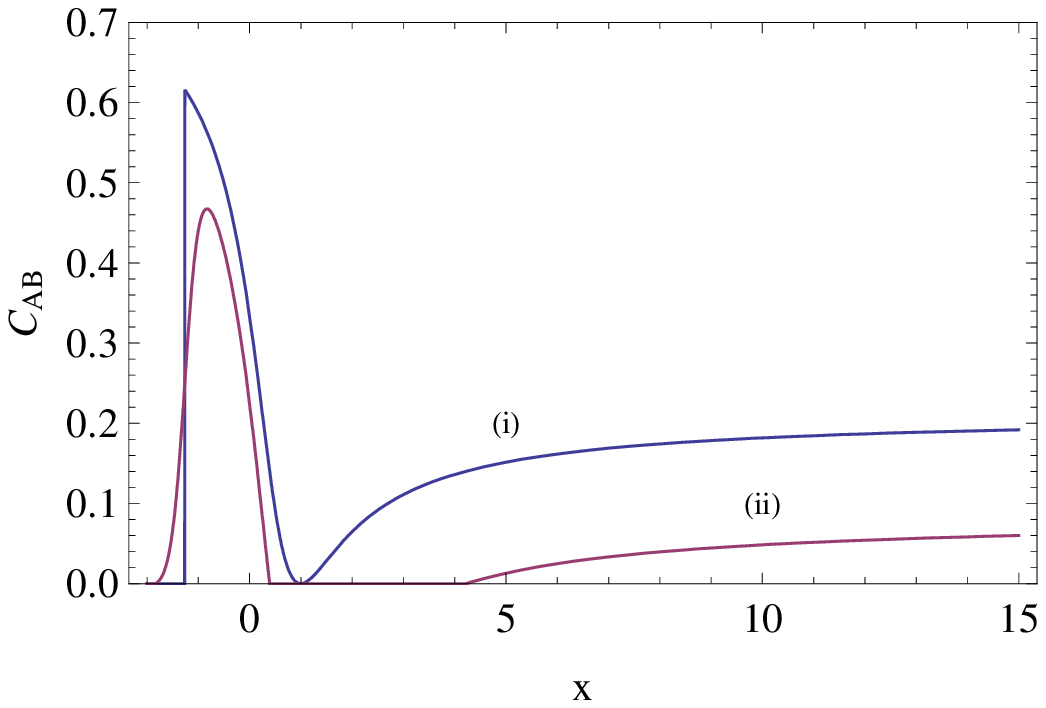}}
\caption{Variation of concurrences (a) $C_{AC}$ and (b) $C_{AB}$ versus $x$ at different 
temperatures $T$ with $y = 0.5$. The different temperature values are \textbf{(a)} 
(i) $T=0$, (ii) $T=0.5$, (iii) $T=2$ and \textbf{(b)} (i) $T=0$, (ii) $T=0.5$}
\end{center}
\end{figure}
\noindent Figures 11(a) and 11(b) show the plots of the nn and nnn entanglements, $C_{AC}$ and 
$C_{AB}$ respectively, versus $x$ for $y=0.5$ and at different values of $T$. 
Figure 12(a) shows the plot of $C_{AC}$ versus $T$ for different negative values of $x$ with 
$y=0.5$. Similar plots are obtained in other ranges of $x$ values. Figure 12(b) shows how 
$C_{AB}$ varies as a function of $T$. 

We lastly calculate the entanglement gap temperature \cite{key-17,key-18}, $T_{E}$, as a function of $x$ 
for both zero and non-zero $y$. $T_{E}$ is determined from the relation $U(T_{E})=E_{sep}$, 
where $U(T)\left(=-\frac{1}{Z}\frac{\partial Z}{\partial\beta}\right)$ is the thermal 
energy at temperature $T$ and $E_{sep}$ is the ground state energy of the classical spin 
model corresponding to the three-qubit Hamiltonian in equation (5). $E_{sep}$ can be easily 
calculated, e.g., $E_{sep}=\frac{1}{2}\left(-1-2x-y\right)$ for $x\geq 0$. 
For temperature $T<T_{E}$, the thermal state is entangled. Figure 13 exhibits 
the variations  of $T_{E}$ versus $x$ for different values of $y$. 
\begin{figure}
\begin{center}
\subfloat[]{\includegraphics[scale=0.7]{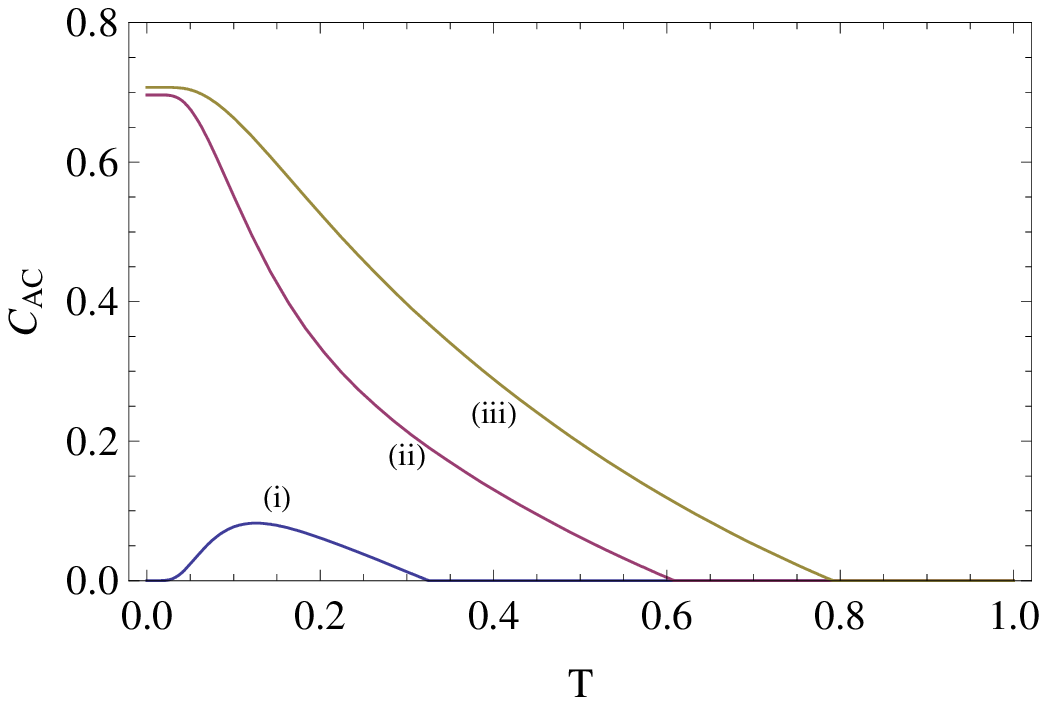}}
\smallskip{}
\subfloat[]{\includegraphics[scale=0.7]{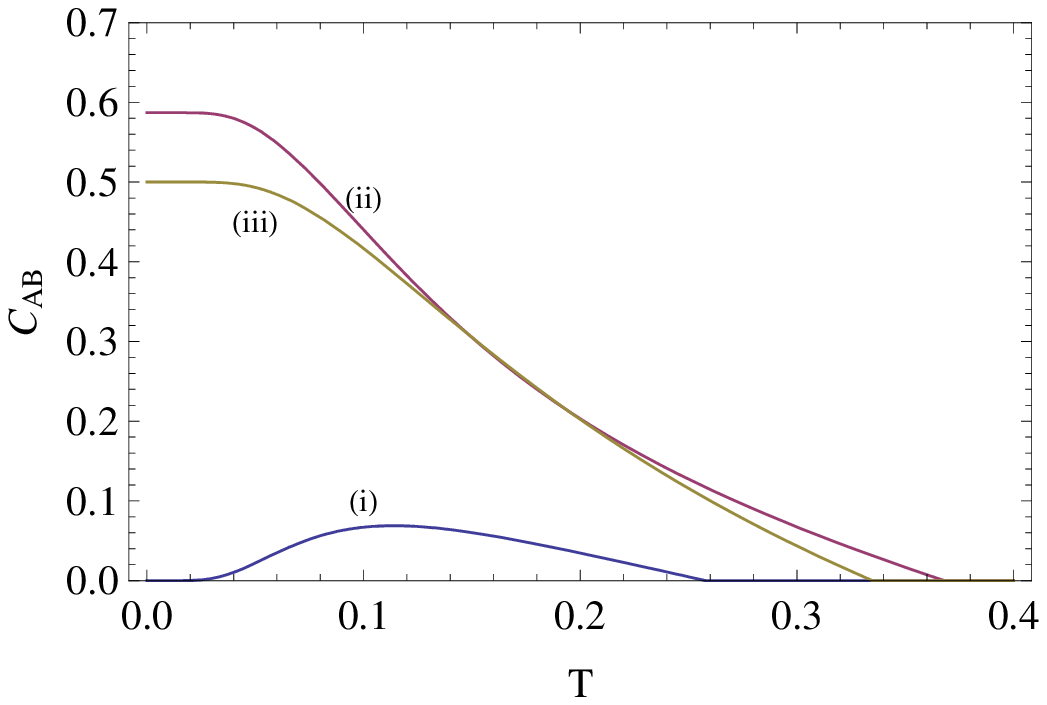}}
\caption{Variation of concurrences (a) $C_{AC}$ and (b) $C_{AB}$ versus $T$ for 
different negative values of $x$ with $y = 0.5$. The different values of $x$ are 
(i) $x=-1.5$, (ii) $x=-1$, and (iii) $x=-0.5$ for both (a) and (b).}
\end{center}
\end{figure}
\begin{figure}
\begin{center}
\includegraphics[scale=0.7]{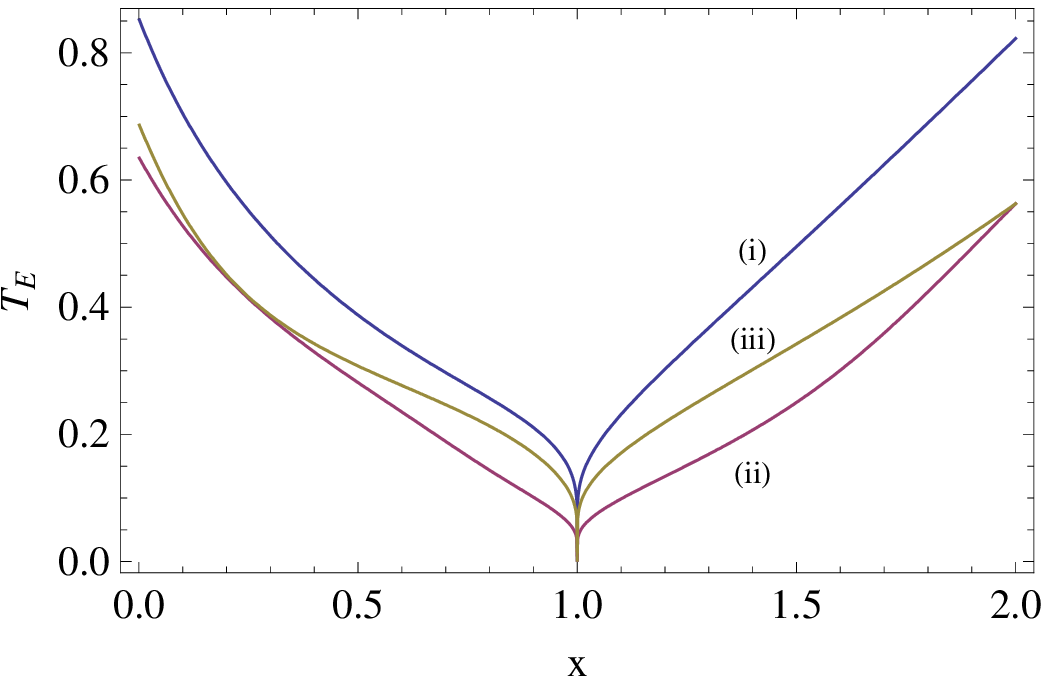}
\caption{Variation of the entanglement gap temperature, $T_{E}$ , versus $x$ for different
magnetic fields. The different values of $y$ are (i) $y=0$, (ii) $y=0.5$, and (iii) $y=1$}
\end{center}
\end{figure}


\section{Summary and Discussion} 
In this paper, we have obtained quantitative measures of pairwise entanglement in a 
molecular three-qubit system as a function of two parameters $x$ and $y$. The 
system represents the recently engineered $Cr_{7}Ni$-$Cu^{2+}$-$Cr_{7}Ni$ complex 
consisting of two $Cr_{7}Ni$ rings coupled via a central $Cu^{2+}$ ion. The parameters 
$x$ and $y$ appearing in the qubit Hamiltonian (equation (9)) have their origins in an effective 
$Cu$-ring axial exchange due to the projection of the rings' dipolar and crystal-field 
anisotropies and an external magnetic field respectively. Timco \textit{et al} \cite{key-3} have 
provided an experimental demonstration that the coupling between molecular spin clusters 
can be manipulated by altering the nature of the linker ions. This opens up the possibility
of chemically controlling the generation of entanglement in spin systems. The molecular 
three-qubit system studied in this paper belongs to a family of clusters with AFM exchange
interactions between the nn ions and a spin-$\frac{1}{2}$ ground state. The simplest case is 
that of a finite chain with an odd number of $S=\frac{1}{2}$ spins and dominant AFM 
interactions between the nn spins. An alternative way of obtaining an $S=\frac{1}{2}$ ground 
state is to replace a single spin in an AFM chain, containing an even number of spins, 
by a spin of different magnitude such that the ground state spin is of magnitude 
$\frac{1}{2}$. The $Cr_{7}Ni$ ring provides an example of the latter possibility. 

In general, the arrangement of spins in a chain can be either linear, or cyclic. 
The molecular three-qubit system studied in this paper is a linear-chain complex, whereas 
the three qubit chain studied in \cite{key-8} is cyclic in nature. Cyclic spin chains 
with an odd number of antiferromagnetically coupled spins have degenerate ground states due 
to magnetic frustration. For a three-spin cyclic chain, the anisotropic Heisenberg XXZ model 
has a four-fold degenerate ground state in zero magnetic field \cite{key-8}. In the presence 
of an external magnetic field, the ground state is doubly degenerate. The effective three-qubit 
Hamiltonian (equation (9)) has the form of the anisotropic Heisenberg XXZ Hamiltonian but 
with a linear i.e. an open-ended structure. In this case, the ground state is doubly degenerate  
in zero magnetic field $(y=0)$ and non-degenerate when $y\neq 0$. The cyclic chain has a greater 
ground state degeneracy because of frustration.

There are prominent differences in the entanglement features of cyclic and linear spin chains. 
As shown in \cite{key-8}, in the case of the AFM cyclic XXZ model, there is no pairwise 
entanglement, as measured by concurrence, for all values of the anisotropy constant. At $T=0$ 
also, the 
concurrence is zero for the AFM case. In contrast, the AFM linear chain has pairwise 
entanglement at both $T=0$ and $T\neq 0$ (figures 3 and 8). One now distinguishes between 
nn and nnn entanglements. In the AFM case, the nnn concurrence $C_{AB}$ is zero at zero 
and finite temperatures whereas the nn concurrence $C_{AC}(=C_{BC})$ is 
non-zero at both $T=0$ and $T\neq 0$. Comparing figures 3 and 6, one finds that on inclusion of 
the magnetic field the range of $x$ values for which $C_{AB}$ (the nnn concurrence) is 
$\neq 0$ is considerably extended. For a 
specific value of $x$, there is, however, a critical value of $y_{c}$ of $y$ such that the 
pairwise entanglement vanishes when $y>y_{c}$. Earlier studies \cite{key-8,key-10} have 
shown that the entanglement between two spins in an AFM chain can be increased 
by raising the temperature or the external magnetic field in specific ranges. This is true for our 
three-qubit system also. In figures 12(a) and 12(b), the curve (i) shows the increase of both 
$C_{AC}$ and $C_{AB}$ with temperature $T$. We have further shown that only pairwise entanglement
exists in the ground state with $y\neq 0$, i.e., there is no three-qubit entanglement as exists 
in the GHZ state (equation (6)).   
One interesting feature of the linear three-spin chain relates to the variation of the threshold 
entanglement temperatures $T_{C}^{(1)}$ and $T_{C}^{(2)}$ versus $x$ for different values of $y$. 
As shown in figure 10(a), the $T_{C}^{(1)}$ versus $x$ plot depends weakly on the values of $y$. The 
threshold temperature $T_{C}^{(2)}$, for nnn entanglement, however, varies more prominently 
with $y$. In the case of the cyclic chain, the single threshold temperature 
depends on both $x$ and $y$. As shown in figure 13, the plots of the entanglement gap 
temperature, $T_{E}$, versus $x$ are different for different values of $y$. In fact, $T_{E}$ has 
a non-monotonic dependence on the values of $y$ (the $y=1$ curve lies in between the $y=0.1$ 
and $y=0.5$ curves). One further notes, 
from figures 10 and 13, that the entanglement gap and threshold temperatures are different for 
the same values of the parameters $x$ and $y$. In fact, one finds that 
$T_{C}^{(2)}<T_{E}<T_{C}^{(1)}$. Figures 4-7 and Figures 11-12 have been obtained by fixing either 
$x$ or $y$ at a specific value. The observations are, however, general in nature and hold true 
in extended ranges of $x$ and $y$ values. In the model studied by us, we have assumed that the 
gyromagnetic factors $g_{A}$, $g_{B}$, $g_{C}$ are of equal amplitude $g$ (equation (8)). In the 
case of the engineered three-qubit system, the diagonal tensors $g_{A,B}$ and $g_{C}$ are 
different. Assuming $g_{A}=g_{B} \ne g_{C}$, ($g_{C(zz)}=2.07$, $g_{A,B(zz)}=1.79$, as quoted in 
\cite{key-3}), we find no qualitative changes in the results reported in sections 2 and 3. It will 
be of interest to study the general case of the magnetic field pointing in an arbitrary 
direction.

The three-qubit molecular cluster exhibits first-order QPTs at specific values of $x$ and $y$. 
In figure 3(a), the QPT at $x=-2$ separates two phases, for $x<-2$ the ground state has no 
entanglement whereas for $-2<x<1$, the ground state, described by the mixed state in equation (24), 
has pairwise entanglement. Similarly, as shown in figure 7, a first order QPT 
occurs at $y=y_{c}$. The point $x=1$ is of special interest as the ground 
and thermal states become separable at this point. The threshold entanglement temperatures, 
$T_{C}^{(1)}$ and $T_{C}^{(2)}$, drop sharply to zero at $x=1$. The first-order transition points can 
be shifted by changing the parameters $x$ and $y$. For example, the transition point $x_{c}$
can be shifted towards higher values by increasing $y$. The first order QPTs are marked by 
discontinuities in the magnitude of both the nn and nnn concurrences associated with the ground 
states. The molecular three-qubit system, $Cr_{7}Ni$-$Cu^{2+}$-$Cr_{7}Ni$, has been specifically
engineered with QIP applications in mind. Since entanglement is a fundamental resource in such 
applications, a knowledge of its dependence on the relevant parameters of the system will be of 
use in the designing and implementation of QIP protocols. With possibilities for controlling the 
couplings in molecular qubit systems \cite{key-3} and realizations of spin Hamiltonians in 
optical lattices \cite{key-19}, some of the theoretical results could be observed in actual 
experiments.
      
\smallskip{}
\noindent \textbf{Acknowledgement :} The Authors thank Amit Tribedi for some useful discussions.



\end{document}